\documentclass[12pt]{article}
\usepackage{amsfonts}
\usepackage{graphicx}
\usepackage{amsthm}
\usepackage{amsmath}
\usepackage{fancyhdr}
\usepackage{enumerate}
\usepackage[agsm]{harvard}
\citationmode{abbr}
\title{Simulation-based Bayesian analysis for multiple changepoints}
\author{Jason Wyse and Nial Friel \\ University College Dublin, Belfield, Dublin 4, Ireland\\ \texttt{jason.wyse@ucd.ie, nial.friel@ucd.ie}}
\date{November, 2010}

\begin{document}
\maketitle

\newcommand{\ctau}{\hbox{\it$\tau$}}
\newcommand{\cov}{\mbox{cov}}
\newcommand{\diag}{\mbox{diag}}
\newcommand{\Var}{\mbox{var}}
\newcommand{\otherwise}{\mbox{otherwise}}
\newcommand{\boldtheta}{\mbox{\boldmath{$\theta$}}}
\newcommand{\boldlambda}{\mbox{\boldmath{$\lambda$}}}
\newcommand{\boldLambda}{\mbox{\boldmath{$\Lambda$}}}
\newcommand{\boldbeta}{\mbox{\boldmath{$\beta$}}}
\newcommand{\boldmu}{\mbox{\boldmath{$\mu$}}}
\newcommand{\boldvarepsilon}{\mbox{\boldmath{$\varepsilon$}}}
\newcommand{\boldOmega}{\mbox{\boldmath{$\Omega$}}}
\newcommand{\boldSigma}{\mbox{\boldmath{$\Sigma$}}}
\newcommand{\boldrho}{\mbox{\boldmath{$\rho$}}}
\newcommand{\boldphi}{\mbox{\boldmath{$\phi$}}}
\newcommand{\mle}{\mbox{m.l.e.}}
\newcommand{\iid}{\mbox{i.i.d.}}
\newcommand{\Like}{\mathcal{L}}
\newcommand{\normal}{\mbox{N}}
\newcommand{\by}{\mathbf{y}}
\newcommand{\be}{\mathbf{e}}
\newcommand{\bu}{\mathbf{u}}
\newcommand{\bX}{\mathbf{X}}
\newcommand{\bx}{\mathbf{x}}
\newcommand{\bV}{\mathbf{V}}
\newcommand{\bz}{\mathbf{z}}
\newcommand{\VO}{$\dot{\mbox{V}}\mbox{O}_2 $ \hspace{0.1mm}}
\newcommand{\VCO}{$\dot{\mbox{V}}\mbox{CO}_2$ \hspace{0.1mm}}
%
%%\fancyhead[L]{Multiple changepoint problems}
%
%
%\LARGE
%\noindent {Bayesian analysis of multiple changepoint problems using MCMC}
%\vspace{1cm}
%
%\large
%\noindent Jason Wyse and Nial Friel \\ University College Dublin, \\Belfield, \\Dublin 4, \\Ireland\
%\vspace{1cm}
%
%\noindent {\bf Running title}\\ Bayesian analysis of multiple changepoint problems\\
%
%\noindent {\bf Corresponding author}\\ Jason Wyse\\
%
%\noindent{\bf Address} \\School of Mathematical Sciences (Statistics),
%\\ University College Dublin,\\
%Belfield, Dublin 4 \\
%Ireland\\
%
%
%\noindent{\bf Email address}\\
%jason.wyse@ucd.ie\\
%
%
%\noindent{\bf Telephone}\\
%+353 1 716 7112 \\
%
%
%\noindent{\bf Facsimile}\\
%+353 1 716 1186\\
%
%\newpage
%
{\bf Abstract:} This paper presents a Markov chain Monte Carlo method to generate approximate posterior samples in retrospective multiple changepoint problems where the number of changes is not known in advance. The method uses conjugate models whereby the marginal likelihood for the data between consecutive changepoints is tractable. Inclusion of hyperpriors gives a near automatic algorithm providing a robust alternative to popular filtering recursions approaches in cases which may be sensitive to prior information. Three real examples are used to demonstrate the proposed approach.
%\vspace{1cm}
%
\newline 

\noindent{\bf Keywords:} Bayes factor;  changepoint; marginal likelihood; model search.
%\chapter{Changepoint sampler}
%
%\section{Models \& methodology} \label{sec_mod_and_meth}
%
%\subsection{Changepoint models}
%
%Consider data which is time ordered and whose underlying generating process undergoes changes during the collection of the data. A changepoint model is formulated by observing that data between changepoints exhibits homogeneity, but globally data exhibits heterogeneity. We call the data between any two changepoints a {\it segment}. We restrict ourselves to modelling data in each segment using the same type of model (see~\citeasnoun{Fearnhead07} for an example of choice of model within segments). Data within segments are assumed independent of data in other segments. So for example if only one changepoint occurs in the data, the model for the entire data should comprise two sub-models; one for the data before the change, and one for the data after, and data before the change is assumed independent of data after.

\section{Introduction}

%The range of applications of changepoint modelling is evident from the substantial volume of literature devoted to this problem in the  Econometrics, Signal Processing and Bioinformatics literatures. One requires a procedure to detect changepoints whenever the underlying process generating data undergoes changes over time and the assumption of homogeneity in the data (in the context of an underlying assumed model) is not appropriate. A reasonable model for such data should account for the heterogeneity of the data exhibited between changes. Examples of such models include Poisson processes in which the intensity undergoes changes~\cite{Raftery86},  changing regression lines~\cite{Carlin92}  and Binomial experiments with changing success probability~\cite{Stephens94}.

The range of applications of changepoint models is evident from the substantial volume of literature devoted to this problem in the  econometrics, signal processing and bioinformatics literatures. A process generating data can often undergo changes over time such that one model will not be appropriate for all time periods. Here ``time'' refers to some natural sequential indexing of the data. Some examples are occurences of coal mining disasters during the $18^{\mbox{th}}$ and $19^{\mbox{th}}$ century~\cite{Raftery86}, DNA or protein composition analysis over base number~\cite{Liu99} and winning streaks in sports~\cite{Yang04}.    

%When the underlying process generating data changes over time the assumption of model homogeneity is not fulfilled. Such data requires a model which allows the underlying parameters to change at the points in time corresponding to changes in the underlying process in order to model heterogeneity over time. A model like this is termed a changepoint model. These models arise frequently in applications to real life data.

%%Bayesian approaches are particularly suited to the analysis of changepoint problems because of their relative simplicity over frequentist approaches. Standard 
%Markov chain Monte Carlo (MCMC) techniques can generally be used to estimate models conditional on a fixed number of changepoints. When the number of changes that occur is unknown, model estimation and inference become more challenging problems.~\citeasnoun{Chib98}, for example, estimates a collection of models, each conditional on a fixed number of changepoints, using MCMC, and then compares these models via marginal likelihood estimated from the MCMC output.~\citeasnoun{Green95} introduces reversible jump MCMC (RJMCMC) with an application to the coal mining disaster data of~\citeasnoun{Jarrett79}, modelling the data as arrivals from a Poisson process with changing intensity, the number of changes and when they occurred both being unknown.

Markov chain Monte Carlo (MCMC) techniques can be used to estimate models with a fixed number of changepoints. When the number of changepoints is unknown, inference is more challenging.~\citeasnoun{Chib98} estimates a collection of changepoint models and compares these using Bayes factors estimated from the MCMC output.~\citeasnoun{Green95} uses reversible jump MCMC (RJMCMC) to explore the number of changepoints in the coal mining disaster data. RJMCMC allows moves between models which satisfy detailed balance.

The use of alternatives to MCMC has grown in this area in recent years. \citeasnoun{Fearnhead06} uses filtering recursions to derive the posterior distribution of changepoints. This can be done for both a known and unknown number of changepoints. An advantage of this approach is that one can draw independent samples from the posterior. MCMC can only do this approximately at best. Extension to online analysis of changepoint models is also possible~\cite{Fearnhead07}. However methods based on filtering recursions rely on strong prior information in most cases. This paper aims to offer an efficient MCMC alternative which can overcome strong reliance on prior assumptions as encountered in recursive computing approaches. The class of models considered is similar to~\citeasnoun{Fearnhead06}. For this reason it is possible that this could be used to give useful starting values for an analysis using filtering recursions. %This is also investigated.

% They may also be readily adapted to online inference for changepoint problems as in~\citeasnoun{Fearnhead07}, something which is not achieved by standard MCMC approaches. However, these methods although powerful, also have potential drawbacks (see Section~\ref{sec:well_log_data}).

%In this paper we demonstrate that the class of conjugate models considered in~\citeasnoun{Fearnhead06} may also be used for efficient MCMC inference for retrospective analysis of multiple changepoint problems.~\citeasnoun**{Punskaya02} use similar models in a RJMCMC implementation to estimate the number of changes and their positions in a segmented regression problem. Our method does not use RJMCMC however, avoiding some of the well documented implementation issues associated with it. 

%The class of models considered is
%
%
%The strategy proceeds by rephrasing the changepoint inference problem as a model choice problem, where a given number of changepoints and their positions corresponds to one particular model. The MCMC method then amounts to a stochastic model search over the large number of possible models, with moves between models with different numbers of changepoints being easily achieved. It may be wise to avoid models which aren't entirely viable, for example, models where there are less than, say, two points between consecutive changepoints. It will be shown that avoiding such models in the search may be easily achieved.

Qualitatively, the work in this paper is similar in some aspects to work by~\citeasnoun{Lavielle01} and~\citeasnoun{Punskaya02} in terms of the class of models considered. The sampling aspect of the approach bears similarities to the samplers of~\citeasnoun{Lavielle01} and~\citeasnoun{Giron07}. This paper extends these works to a broader range of data models and proposes a more efficient way of sampling changepoints. An aim is also to highlight possible shortcomings of alternatives to MCMC and how these could be overcome by using simulation approaches to inform choices for recursive computing approaches.

The remainder of the paper is  organised as follows. In Section~\ref{sec_mod_and_meth} the type of changepoint model under consideration is presented. Section~\ref{sec:collapsing_cp_models} reviews the reversible jump approach to changepoint estimation and discusses how this can be simplified into a fixed dimensional sampling scheme. Section~\ref{sec:sampler} gives the moves to sample from the simpler fixed dimensional posterior. Prior specification is discussed in Section~\ref{sec:priors}, and Section~\ref{sec:recursions} reviews the filtering recursion approach to generating samples of changepoints. Performance of the sampler is validated by analyzing the coal mining disasters data in Section~\ref{sec:coal_mining}, while Sections~\ref{sec:tiger_woods} and~\ref{sec:well_log_data} compare qualitative aspects of the simulation based sampler approach and filtering recursions approach using two real data examples. A brief discussion concludes the article.

%The method is demonstrated using three examples in Section~\ref{sec:examples}. The first of these concerns the issue of ``streakiness'' in sports and examines data collected on Tiger Woods' championships win record from September 1996- June 2001. For the second example we analyze the Coal mining disasters data of~\citeasnoun{Jarrett79}. The last example we consider looks at the Well-log data of~\citeasnoun{ORuanaidh96}. In this example we consider the incorporation of hyperpriors in the models presented, and provide a comparsion of our methods with those of~\citeasnoun{Fearnhead06}. We conclude the paper with a brief discussion.

\section{Changepoint models} \label{sec_mod_and_meth}

Consider the data $y_{1:n} = (y_1,\dots,y_n)$ which is time ordered. Here $y_i$ is observed before $y_j$ if $i<j$. Time in this context can refer to any natural ordering of the data as it is observed. A changepoint occurs at time $t$ if $y_1,\dots,y_t$ are generated differently to $y_{t+1},\dots,y_n$. Referring to $y_{s:r} (s<r)$ as a segment, this says that the segments $y_{1:t}$ and $y_{t+1:n}$ are heterogeneous between but homogeneous within. Parametric changepoint models assign a different parameter for each segment to account for this heterogeneity.

This paper considers multiple changepoints which will be denoted $\tau_1,\dots,\tau_k$. These split the data into $k+1$ segments. The likelihood for segment $j$ has parameter $\theta_j$. Conditional on a segmentation, the data within each segment is assumed independent. It is also assumed that the regime parameters $\theta_j$ are independent. The likelihood of the segmentation $\tau = (\tau_1,\dots,\tau_k)$ is
\[
\prod_{j=1}^{k+1} \prod_{i=\tau_{j-1}+1}^{\tau_j} \pi(y_i|\theta_j)
\]
where for convenience $\tau_0 = 0, \tau_{k+1}=n$. Instead of using $\tau$, segmentations can be labelled with the binary latent vector $z = (z_1,\dots,z_n)$ with $z_t=1$ indicating a changepoint at time $t$ and $z_n = 0$. Independent priors are assumed for each member of $\theta = (\theta_1,\dots,\theta_{k+1})$ with hyperparameter $\gamma$ and there is a prior for the changepoints with hyperparameter $\xi$, given by $\pi(z|k,\xi)$. The posterior may be written
\begin{eqnarray*}
\pi(z,\theta|y,k,\xi,\gamma) & \propto &\pi(z|k,\xi) \pi(\theta|k,\gamma) \pi(y|\theta,z,k)     \\
															& = & \pi(z|\xi)\prod_{j=1}^{k+1} \pi(\theta_j|\gamma) \prod_{i=\tau_{j-1}+1}^{\tau_j} \pi(y_i|\theta_j)
\end{eqnarray*}
where the dependence on the number of changepoints, $k$, is made explicit. A prior $\pi(k)$ may be introduced so that the posterior of interest is the joint posterior of $(k,z,\theta)$,
\begin{equation}
\pi(k,z,\theta|y,\xi,\gamma) \propto \pi(k) \pi(z,\theta|y,k,\xi,\gamma). \label{eq:hierarchical_mod}
\end{equation}
This is a hierarchical changepoint model similar to that used in~\citeasnoun{Green95}.

\section{Collapsing changepoint models} \label{sec:collapsing_cp_models}

It is possible to construct a MCMC scheme to sample the posterior of (\ref{eq:hierarchical_mod}) using RJMCMC~\cite{Green95}. The sampler will explore the product space support of this posterior:
\[
\mathcal{X} = \prod_k \{k\} \times \{\mathcal{Z}_k,\Theta_k|k\}
\] 
where $\mathcal{Z}_k,\Theta_k$ are respectively the sample spaces of $z$ and $\theta$ conditional on $k$ changepoints. A switch in the number of changepoints in the model can be made by a RJ move switching between support subspaces. For the purposes of illustration a straightforward move of this type is now discussed. When proposing a switch from $k$ to $k+1$ changepoints one possibility is to generate a random variable $u \in \mathbb{R}^d$ and form a bijection $f:\Theta_k \times \mathbb{R}^d \rightarrow \Theta_{k+1}$ where $d$ is the dimension of a single $\theta_j$. This bijection gives the parameters for the proposed $k+1$ changepoint model as a function of those for the $k$ changepoint model; $\theta'=(\theta'_1,\dots,\theta'_{k+2}) = f(\theta_1,\dots,\theta_{k+1},u)$. The proposed switch in model is then accepted with probability $\min(1,R)$ where
\[
R = \frac{\pi(k+1,z',\theta'|y,\xi,\gamma) }{\pi(k,z,\theta|y,\xi,\gamma) } \frac{P(k+1,k)}{P(k,k+1)} \frac{1}{ q(u|\theta)} \left|\frac{\partial(\theta')}{\partial(\theta,u)} \right|.
\]
In the expression for $R$, $P(\cdot,\cdot)$ denotes the proposal probability for transitions between different numbers of changepoints, and $q(\cdot|\theta)$ is the proposal density of $u$. The last term on the right is a Jacobian term for the bijection $f$. The reverse move in switching from $k+1$ to $k$ changepoints is accepted with probability $\min(1,R^{-1})$. More elaborate moves between support subspaces are possible which propose changes to the model of more than one dimension or involve stochastic moves in both directions.
%Indeed, one can generate a random variable $u'$ and set the bijection $f$ so that $(\theta',u') = f(\theta,u)$ and the acceptance probability becomes $\min(1,R)$ with 
%\[
%R = \frac{\pi(k+1,z',\theta'|y,\xi,\gamma) }{\pi(k,z,\theta|y,\xi,\gamma) } \frac{P(k+1,k)}{P(k,k+1)} \frac{q'(u'|\theta')}{ q(u|\theta)} \left|\frac{\partial(\theta',u')}{\partial(\theta,u)} \right|.
%\]
%as long as $\dim(\theta') + \dim(u') = \dim(\theta)+\dim(u)$ (dimension matching).

The key questions in a changepoint analysis are usually; how many changepoints are there and where are the changepoints? The segment parameters $\theta$ can be viewed as a nuisance parameter in this regard. Choosing conjugate priors for the $\theta_j$ allows these to be collapsed in the model 
\begin{eqnarray}
\pi(k,z|y,\xi,\gamma) & \propto &\pi(k)\pi(z|k,\xi) \prod_{j=1}^{k+1} \int \pi(\theta_j|\gamma) \prod_{i=\tau_{j-1}+1}^{\tau_j} \pi(y_i|\theta_j)\, \mbox{d} \theta_j \nonumber\\
											& = & \pi(k)\pi(z|k,\xi) \prod_{j=1}^{k+1} \pi(y_{\tau_{j-1}+1:\tau_j}|\gamma), \label{eq:collapsed_mod}
\end{eqnarray}
where $\pi(y_{\tau_{j-1}+1:\tau_j}|\gamma)$ is the marginal likelihood of the data segment $y_{\tau_{j-1}+1:\tau_j}$ and is assumed to be available in closed form due to the conjugacy. The support of this posterior is 
\[
\mathcal{Y} = \prod_k \{k\} \times \{\mathcal{Z}_k|k\}
\]
and a switch from $k$ to $k+1$ changepoints does not require the design of a bijective function between support subspaces. The proposed switch in model is now accepted with Metropolis-Hastings probability $\min(1,A)$ where
\begin{equation}
A = \frac{\pi(k+1,z'|y,\xi,\gamma) }{\pi(k,z|y,\xi,\gamma) } \frac{P(k+1,k)}{P(k,k+1)}. \label{acc_collapsed}
\end{equation}
This idea of collapsing has been used previously in~\citeasnoun{Punskaya02} and~\citeasnoun{Lavielle01} for Gaussian data models.

It can be seen that the first term on the right hand side of the acceptance ratio (\ref{acc_collapsed}) is the Bayes factor for a model with $k+1$ changepoints at positions $z'$ versus a model with $k$ changepoints at positions $z$, assuming all models are equally likely, {\it a priori}. Noting this, it becomes apparent that sampling $k$ and $z$ is equivalent to a model search over large model space. If there can be at most $\bar{k}$ changepoints, then the dimension of this space is $\sum_{k=0}^{\bar{k}} \binom{n-1}{k}$. So searching for up to 5 changepoints in a dataset of length 200 corresponds to a dimension $\sim 2.5 \times 10^9$. In the next section an MCMC scheme to search over these large model spaces, that is, sample from the posterior (\ref{eq:collapsed_mod}), is proposed. 

\section{Sampling changepoints}\label{sec:sampler}

The MCMC scheme to generate samples of changepoints from the posterior (\ref{eq:collapsed_mod}) consists of three possible moves: add a changepoint; delete a changepoint; move a changepoint. Each sweep consists of the following; 
\begin{enumerate}[i.]
\item Choose to add or delete a changepoint with probabilities $a_k$ and $d_k = 1-a_k$ respectively. Clearly $a_{\bar{k}} = d_0 = 0$.
\item Select a changepoint and propose to move it to a position in the range of its closest neighbouring changepoints.  
\end{enumerate}

\noindent{\bf Add or delete a changepoint}

\noindent This move has been dicussed in Section~\ref{sec:collapsing_cp_models} but more details are given here. Suppose there is currently $k$ changepoints at postions $z$. Let $z$ correspond to changepoints at $\tau_1,\dots,\tau_k$. Randomly select one of the $n-k-1$ points where there could be a changepoint $\mbox{i.e.}$ a $t < n$ with $z_t = 0$. Say this is currently in segment $j$ given by $y_{\tau_{j-1}+1:\tau_j}$. Relabel the proposed changepoints in $z'$ as $\tau'_1,\dots,\tau'_{k+1}$ with $\tau'_j = t$. Cancellation of marginal likelihood terms then implies that
\[
\frac{\pi(k+1,z'|y,\xi,\gamma) }{\pi(k,z|y,\xi,\gamma) } = \frac{\pi(k+1)}{\pi(k)}\frac{\pi(z'|k+1,\xi)}{\pi(z|k,\xi)} \frac{\pi(y_{\tau'_{j-1}+1:\tau'_j}|\gamma)\pi(y_{\tau'_j+1:\tau'_{j+1}}|\gamma)}{\pi(y_{\tau_{j-1}+1:\tau_j}|\gamma)}
\]
so calculation of $A$ in (\ref{acc_collapsed}) only requires at most three marginal likelihood values. Conversely, for the delete move, one of the $k+1$ changepoints in $z'$ is chosen at random and the calculation of the acceptance probability involves
\[
\frac{\pi(k,z|y,\xi,\gamma) }{\pi(k+1,z'|y,\xi,\gamma) } = \frac{\pi(k)}{\pi(k+1)}\frac{\pi(z|k,\xi)}{\pi(z'|k+1,\xi)} \frac{\pi(y_{\tau_{j-1}+1:\tau_j}|\gamma)}{\pi(y_{\tau'_{j-1}+1:\tau'_j}|\gamma)\pi(y_{\tau'_j+1:\tau'_{j+1}}|\gamma)}.
\]
Finally, the proposal one step transition probabilities for the number of changepoints will be $P(k,k+1) = a_k/(n-k-1)$ and $P(k+1,k) = d_{k+1}/(k+1)$, so that $A$ (\ref{acc_collapsed}) can be computed. The acceptance probability for the add move is then $\min(1,A)$ and the delete move is accepted with probability $\min(1,A^{-1})$.
\newline

\noindent{\bf Move a changepoint}\label{sec:move_changepoint}

\noindent{\it Gibbs update:} Given the model assumption that the marginal likelihood for any segment is available in closed form, it is possible to update the position of any changepoint from its full conditional. Suppose $\tau_j$ is being updated. Then the conditional probability that $\tau_j=t$, $\tau_{j-1} < t < \tau_{j+1}$ is proportional to
\[
\pi(z'_{(t)}|k)\pi(y_{\tau_{j-1}+1:t}|\gamma)\pi(y_{t+1:\tau_{j+1}}|\gamma)
\]
where $z'_{(t)}$ corresponds to changepoints $\tau_1,\dots,\tau_{j-1},t,\tau_{j+1},\dots,\tau_k$. The effort required for the Gibbs update is $O(\tau_{j+1} - \tau_{j-1})$ and so may be computationally expensive for large datasets with changepoints far apart, or datasets with many changepoints. In this situation a local random walk update may be preferred.

%Another interesting move is to choose a changepoint $\tau_l$ at random and propose to place it in segment $y_{\tau_{j-1}+1:\tau_j}$, where $|j-l|>1$. The proposed position is drawn from the Gibbs distribution above. The Gibbs probabilities are included in the Metropolis-Hastings acceptance ratio. 

\noindent{\it Local random walk update:} $t$ is drawn uniformly from the integers $\max(\tau_j - l,\tau_{j-1}+1), \dots,\min(\tau_j + l,\tau_{j+1}-1)$ where $l$ specifies the locality of the proposed move. The move is accepted with probability $\min(1,B)$ where
\[
B = \frac{\pi(y_{\tau_{j-1}+1:t}|\gamma)\pi(y_{t+1:\tau_{j+1}}|\gamma)}{\pi(y_{\tau_{j-1}+1:\tau_j}|\gamma)\pi(y_{\tau_{j}+1:\tau_{j+1}}|\gamma)}.
\] 
In the event that $\tau_j-l \leq \tau_{j-1}$ and $t < \tau_j$, $B$ must be multiplied by $(\tau_j-\tau_{j-1}+l)/(t - \tau_{j-1}+l)$. Similar modifications are needed if $t>\tau_j$ or $\tau_{j}+l \geq \tau_{j+1}$.
%$\tau_{j-1}+m+1, \dots, \tau_j-m$

\noindent{\it Mixture of updates:} A mixture of the two moves above should improve mixing and not be overly computationally expensive. For example, choose the Gibbs update with probability $g_k=1/\sqrt{k}$ $(k \geq 1)$ and random walk with probability $r_k = 1-g_k$.

\section{Prior specification} \label{sec:priors}

There are many possible choices for $\pi(z|k,\xi)$.~\citeasnoun{Yao84} considers a geometric distribution for the duration, $d$, of segments; $d \sim \mbox{Geometric}(p)$. The prior used by~\citeasnoun{Green95} has been adapted by~\citeasnoun{Fearnhead06} for the discrete time context discussed here. The $k$ changepoint locations are distributed as the even numbered order statistics in a sample of size $2k+1$ from the integers $1,\dots,n-1$, drawn without replacement. 

The geometric prior relies on specification of $\xi = p$. Ideally, one could simulate a segment specific $p_j$ in a similar vein to~\citeasnoun{Chib98}. However this leads to more difficult jump dynamics when adding or deleting a changepoint. The choice of $p$ may impact the analysis. If too small, then it will assign very small probability to changepoints, meaning small changes cannot be detected with high power. If too large, then spurious changepoints are inferred. For these reasons, it desireable to introduce a hyperprior on $p$. For example, a $\mbox{Beta}(\alpha_1,\alpha_2)$ prior with $1<\alpha_1<\alpha_2$ (more weight less than 0.5), would be an ideal choice if there is enough prior information to choose $\alpha_1,\alpha_2$. Otherwise, a non-informative $\mbox{Beta}(1,1)$ prior would suffice.  

%A prior which gives zero weight to segments less than a certain duration is also considered here. Suppose there is prior knowledge that segments will have duration at least $r$. Then changepoints can be distributed uniformly on intervals whose endpoints are $r$ points from the closest changepoint. 

Segment parameters share a common hyperparameter $\gamma$ in Section~\ref{sec_mod_and_meth}. It is therefore possible to explore uncertainty in $\gamma$ also by introducing a hyperprior $\pi(\gamma)$.

Sampling $p$ and $\gamma$ can be easily incorporated into the MCMC scheme in Section~\ref{sec:collapsing_cp_models}. One sweep of the algorithm consists of:
\begin{enumerate}
\item Sample the changepoints.
\item Conditional on the changepoints sample $p$.
\item Conditional on the changepoints sample $\theta$.
\item Conditional on $\theta$ sample $\gamma$ and discard the $\theta$ values.
\end{enumerate}
For the last step here, it will often be possible to sample $\gamma$ using a Gibbs step. However, if this is not possible, a simple random walk Metropolis-Hastings could be used. 

\section{Analysis by filtering recursions} \label{sec:recursions}

It is useful to give a brief recap of the filtering recursions analysis of~\citeasnoun{Fearnhead06} based on a point process prior for changepoint positions. \citeasnoun{Liu99},~\citeasnoun{Barry92} have also used these types of methods for the analysis of changepoint problems. Define
\[
R_\gamma (t) = \Pr\{y_{t:n}|\mbox{changepoint at }t-1, \gamma\}.
\]
It is possible to compute this quantity in a backward recursion. Defining $R_\gamma (n) = \pi(y_n|\gamma)$, for $t=n-1,\dots,2$
\[
R_\gamma(t) = \sum_{s=t}^n \pi(y_{t:s}|\gamma)R_\gamma (s+1)g(s-t+1) + \pi(y_{t:n}|\gamma) (1-G(n-t+1))
\]
and 
\[
R_\gamma(1) = \sum_{s=1}^{n-1} \pi(y_{1:s}|\gamma)R_\gamma (s+1)g_0(s) + \pi(y_{1:n}|\gamma)(1-G_0(n-1))
\]
where the dependence of $R_\gamma (t)$ on the hyperparameter $\gamma$ has been made explicit. Here $g(\cdot)$ gives the point process for the changepoint positions and $G(\cdot)$ the corresponding cumulative distribution function (the subscript $0$ on $g$ and $G$ in $R_\gamma(1)$ denotes the distribution of the first changepoint after $0$).~\citeasnoun{Yao84} takes this as geometric as do~\citeasnoun{Barry92}.~\citeasnoun{Fearnhead06} suggests a negative binomial family in general for this process. 

After computing the recursions, a sample of size $N$ of the changepoints can be efficiently simulated as follows:
\begin{enumerate}
\item Initialize all samples to have a changepoint at $t=0$.
\item For $t=0,\dots,n-2$
\begin{enumerate}
\item Get $n_{t}$, the number of samples for which the last changepoint was at time $t$.
\item If $n_t>0$ compute the distribution of the next changepoint:
\[\Pr\{\tau|y_{1:n},t\} = \pi(y_{t+1:\tau}|\gamma)R_{\gamma}(\tau+1)g(\tau-t)/R_{\gamma}(t+1)\]
\item Sample $n_t$ times from $\Pr\{\tau|y_{1:n},t\}$ and update the $n_t$ samples that have the last changepoint at $t$.
\end{enumerate}
\end{enumerate}
There are two strengths of this approach. The first is that the samples of changepoints will be independent draws from the posterior distribution. The second is the fast sampling algorithm which avoids computing the distribution of the next changepoint for each possible time. The main weakness of this approach is that the generated samples are dependent on a fixed value of the hyperparameters $\gamma$. Updating $\gamma$ using a hyperprior to correctly explore uncertainty in the value would involve recomputing the recursions $R_{\gamma}(t)$ for each new value of $\gamma$, a computation which is quadratic in $n$. This would lead to an infeasible computational overhead for any reasonably large sample from the posterior.

\section{Poisson data: coal mining disasters}\label{sec:coal_mining}

\begin{figure}
\begin{center}
$\begin{array}{cc}
\includegraphics[width=60mm]{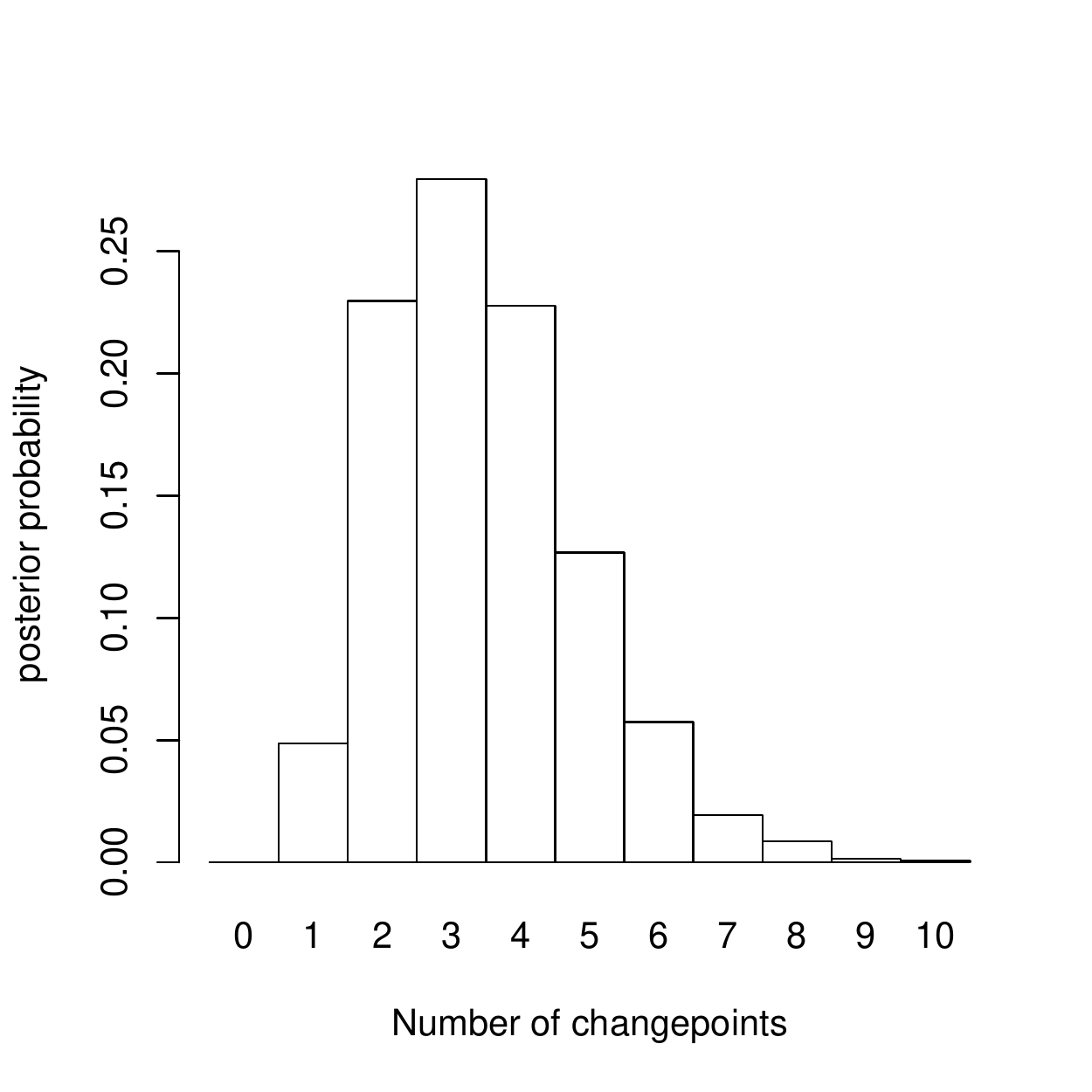}&
\includegraphics[width=60mm]{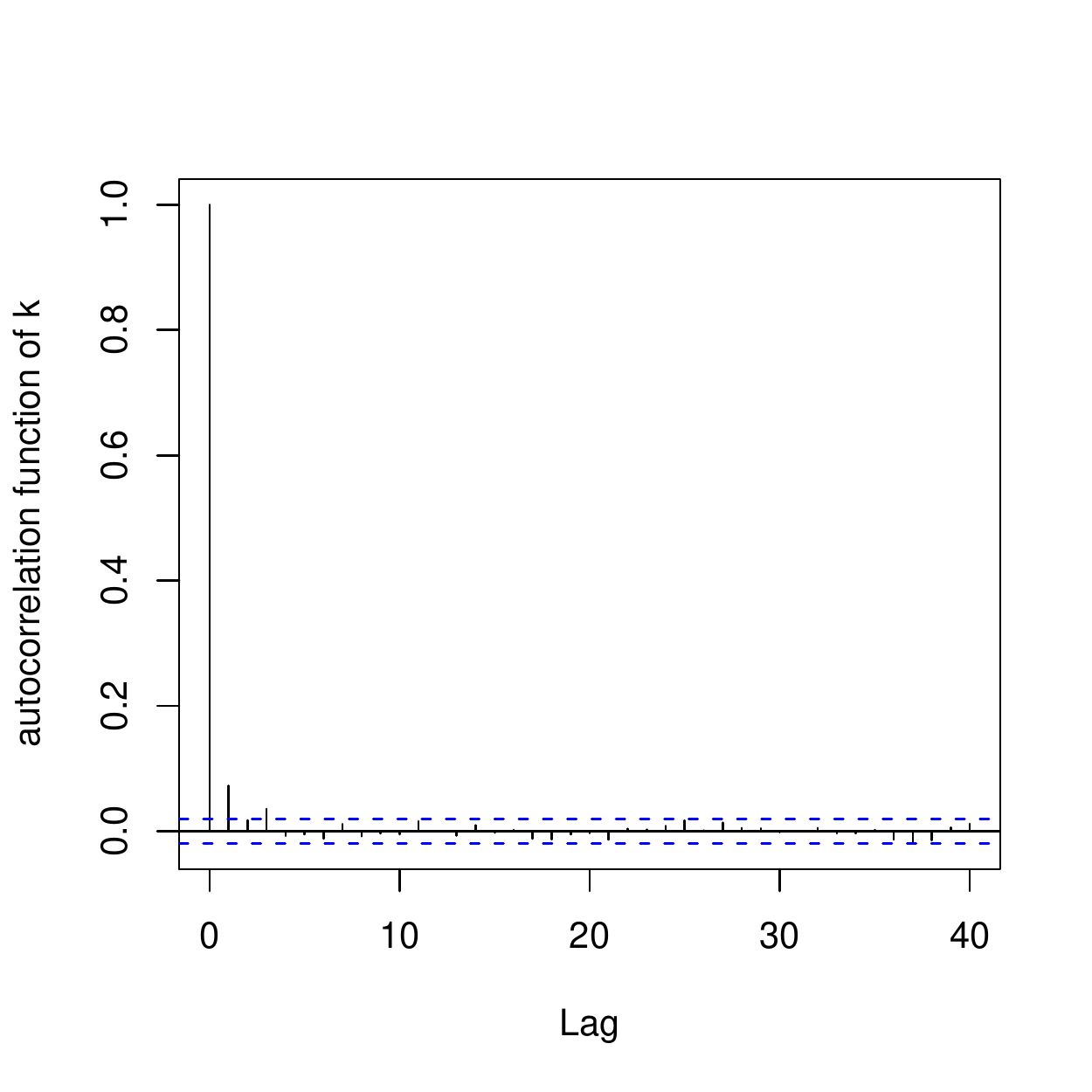}\\
\mbox{(a)} & \mbox{(b)} 
\end{array}$
\end{center} \caption{Coal mining disasters: (a) Posterior number of changepoints (b) Plot of the autocorrelation function of the number of changepoints}\label{fig:Coal_mining_posterior} 
\end{figure}

The sampler of Section~\ref{sec:sampler} was applied to the coal-mining data of~\citeasnoun{Jarrett79}. This data records the dates of serious coal-mining disasters between 1851 and 1962. Disasters are assumed to arise from a Poisson process whose intensity is the height of a step function with an unknown number of steps. For comparison with~\citeasnoun{Fearnhead06}, time is discretized in weeks and the intensities are taken to be $\mbox{Gamma}(1,200/7)$, {\it a priori}. Details on the model marginal likelihood calculations are given in the Appendix. Conditional on $k$ changepoints the prior on their positions was taken to be the same as the distribution of the even numbered order statistics of a sample of size $2k+1$ drawn without replacement from $\{1,\dots,n-1\}$ \cite{Fearnhead06},
\[
\pi(\tau_1,\dots,\tau_k|k) =\binom{n-1}{2k+1} ^{-1}\prod_{j=0}^{k} (\tau_{j+1}-\tau_j-1),
\]
where for convenience, $\tau_0=0$ and $\tau_{k+1}=n$. The algorithm was run for 500,000 sweeps after 10,000 burn in. Every $50^{\mbox{th}}$ sample was taken to reduce dependency in the MCMC iterates. This took 10 seconds on a 2.5GHz processor. Figure~\ref{fig:Coal_mining_posterior} (a) shows that the posterior number of changepoints is almost identical to that obtained from long runs of a RJMCMC sampler and methods based on recursions (see~\citeasnoun{Fearnhead06}, Figure 1$\mbox{.}$(a)). 

\section{Streakiness in sports}\label{sec:tiger_woods}

\begin{figure}
\begin{center}
$\begin{array}{cc}
\includegraphics[width=60mm]{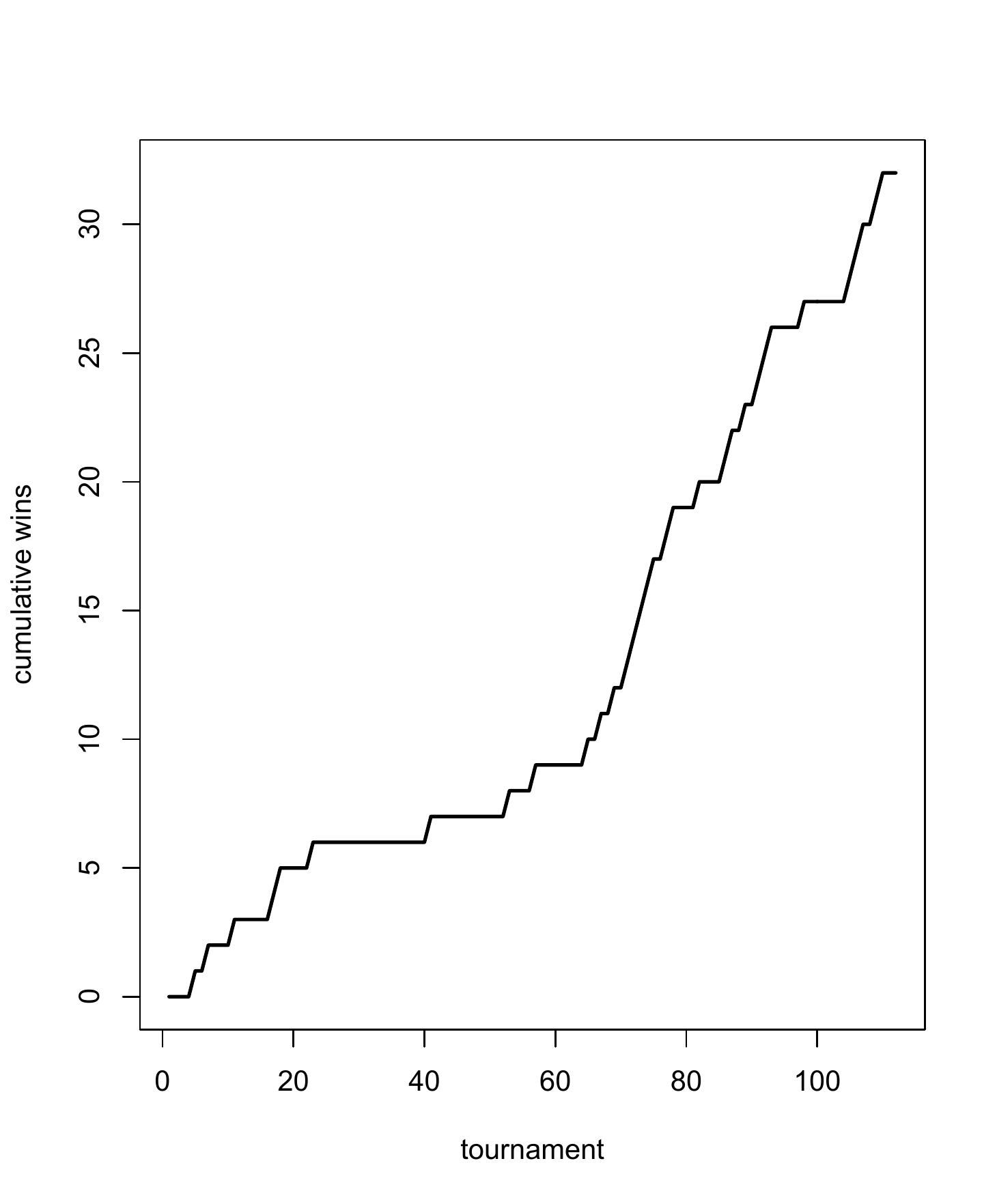} & \includegraphics[width=60mm]{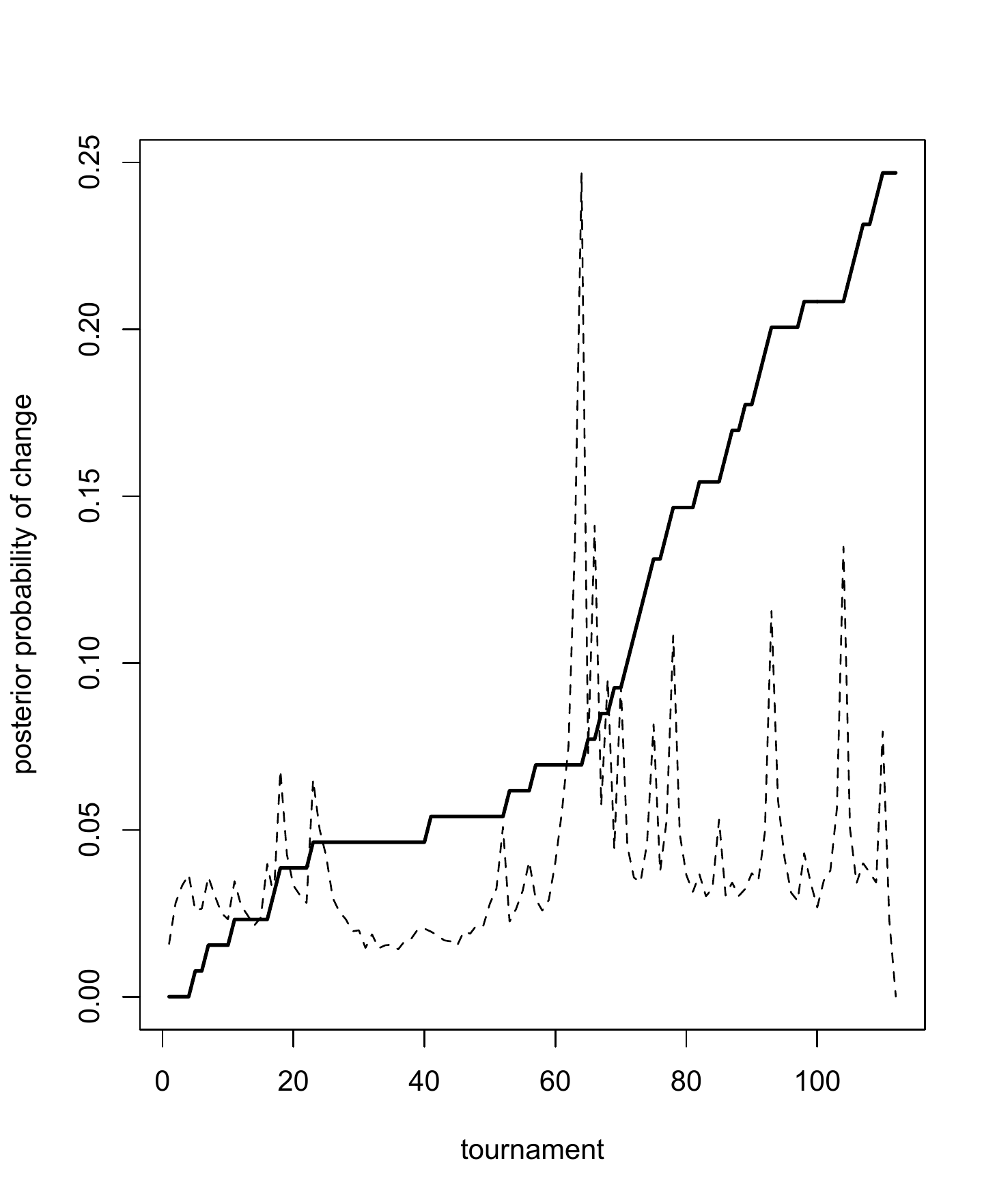}\\
\mbox{(a)}&\mbox{(b)}
\end{array}
$
\end{center}
\caption{Streakiness dataset: Cumulative counts of Tiger Woods' tournament wins}\label{fig:tiger_woods} 
\end{figure}

A sportsperson is considered ``streaky'' if instead of having a constant success rate over time, they have periods of high success rate. Such data will generally be a binary sequence with a ``0'' denoting a loss and a ``1'' denoting a win. The data concerning Tiger Woods' championship wins from September 1996- June 2001 was given and analyzed  by~\citeasnoun{Yang04}, and are reanalyzed using the sampler of Section~\ref{sec:sampler}. The cumulative counts are shown in Figure~\ref{fig:tiger_woods} (a). Following~\citeasnoun{Yang04} the data as is assumed to arise as a sequence of Bernoulli trials, with a possible changing probability of success. The data is ordered by subsequent tournament, and if a changepoint occurs, it is assumed to do so at some tournament. Let %$\theta_j$ denote the probability of success in segment $j$, so that the likelihood of the data in segment $j$ is
%\[
%\pi(\by_{\tau_{j-1}+1:\tau_j}|\theta_j,\bz,k) = \theta_j^{s_j} (1-\theta_j)^{\tau_j-\tau_{j-1}-s_j}
%\]
$s_j = \sum_{i=\tau_{j-1}+1}^{\tau_j} y_i$, the number of sucesses in a segment. Then assuming a $\mbox{Beta}(\alpha,\beta)$ prior for the probability of success in any segment,
\[
\pi(y_{\tau_{j-1}+1:\tau_j}|\alpha,\beta)=\frac{\Gamma\{\alpha+\beta\}}{\Gamma\{\alpha\}\Gamma\{\beta\}}  \frac{\Gamma\{s_j + \alpha\}\Gamma\{\tau_j-\tau_{j-1}-s_j+\beta\}}{\Gamma\{\tau_j-\tau_{j-1}+\alpha+\beta\}}.
\]
Details of this calculation are given in the Appendix. The parameters $\alpha$ and $\beta$ were both set equal to 1. The distribution between changepoints was taken to be $\mbox{Geometric}(p)$. 
%The prior on $k$ was taken to be uniform on the range $0,\dots,10$. This gives no discriminating prior weight on a particular number of changepoints. %
The specification of $p$ may have an effect on the outcome of the analysis. It is thus desirable to investigate uncertainty in its value. This is done in two ways. Firstly, a simulation study using the sampler of Section~\ref{sec:sampler} is carried out, where there is a hyperprior placed on $p$. Secondly, outputs of analyses using filtering recursions~\cite{Fearnhead06} for a range of values $p$ are compared. 

\begin{figure}
\begin{center}
$
\begin{array}{cc}

\includegraphics[width=50mm,height=65mm]{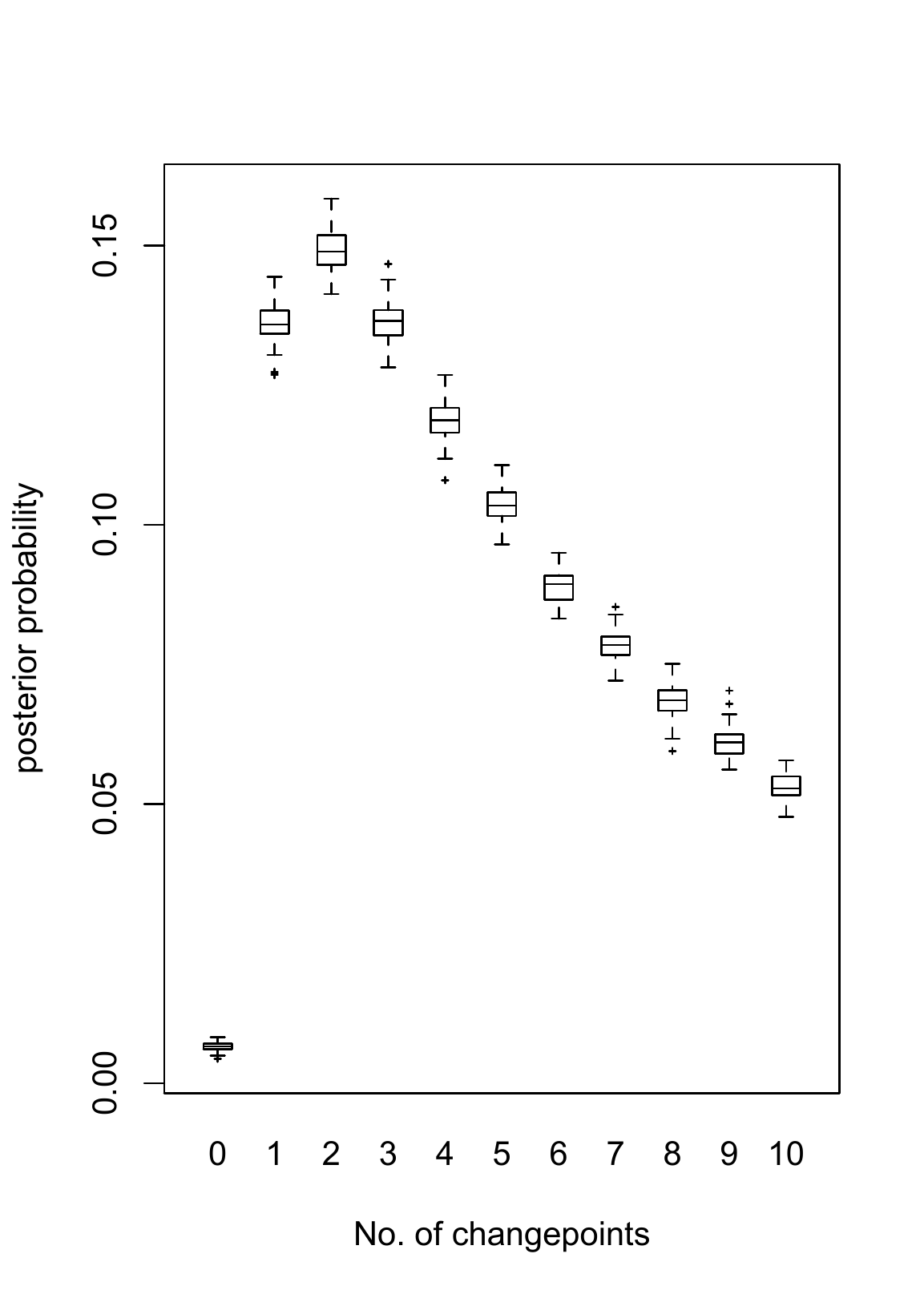}&
\includegraphics[width=60mm,height=65mm]{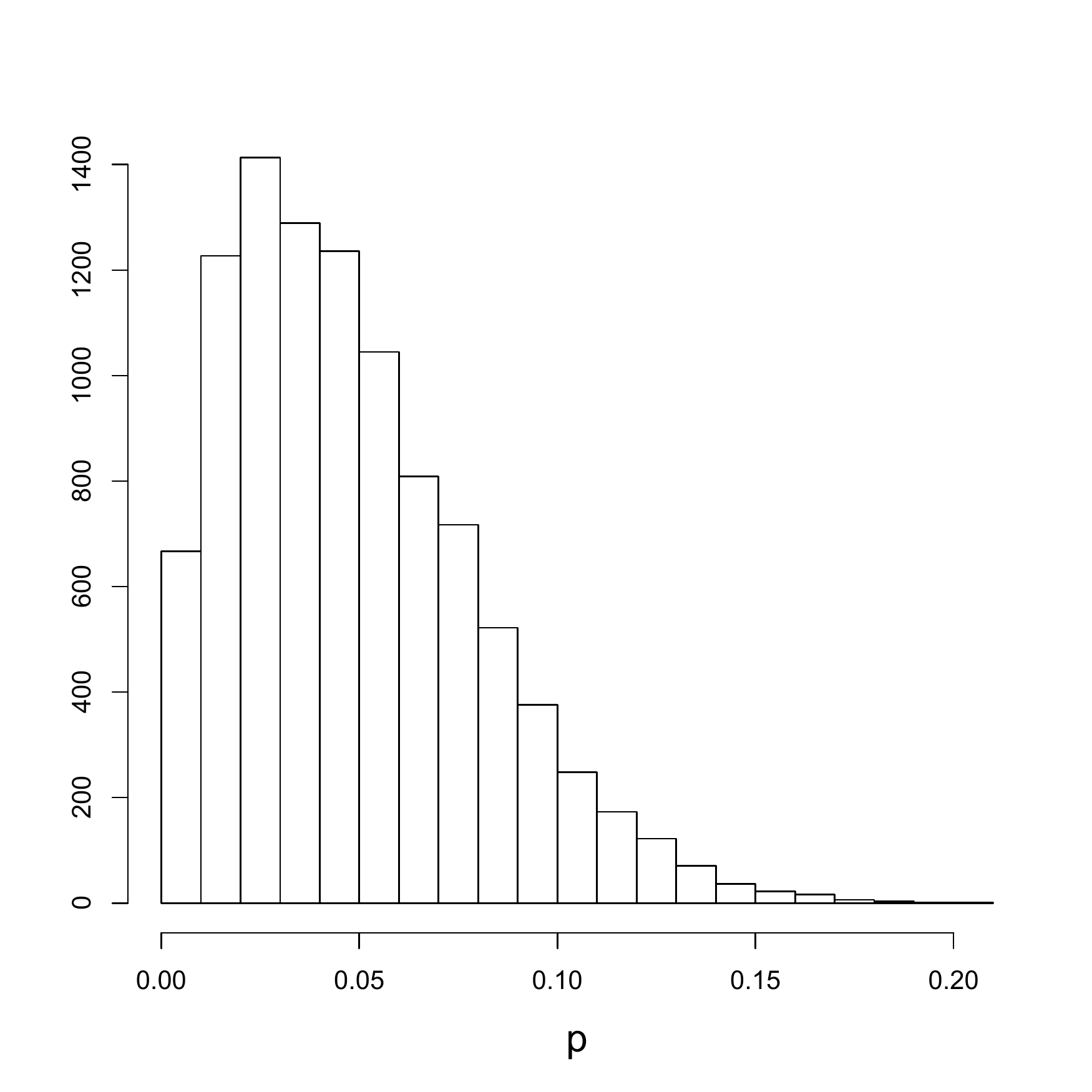}
\\
\mbox{(a)} & \mbox{(b)} 
\end{array}
$
\end{center}
\caption{Streakiness dataset: (a) Boxplots of posterior probability for a given number of changepoints for 100 independent runs of the sampler.(b) Histogram of marginal draws of $p$ from one run in the MCMC sampler simulation study.} \label{fig:tiger_figure}
\end{figure}

%\noindent{\it Simulation study}
For the MCMC simulation study using the sampler proposed earlier, the hyperparameter given to $p$ was uniform on $[0,1]$. After each update of the changepoints the value of $p$ was updated by drawing from its full conditional distribution which is $\mbox{Beta}(k+1,n-k)$. A discrete uniform prior on $[0,\dots,10]$ was taken for the number of changepoints. This gives no discriminating prior weight on a particular number of changepoints. The sampler was run 100 times each for 100,000 burn in iterations and a subsequent 1,000,000 iterations. To reduce dependency in the sample, only every $100^{\mbox{th}}$ sample was stored. Each run took about $1.5$ min on a 2.5GHz processor. Changepoints were updated using the mixture of moves discussed in Section~\ref{sec:move_changepoint}. Figure~\ref{fig:tiger_woods} (b) shows the output from one of these runs, with the posterior probability of a changepoint at any tournament indicated by the dashed line and a scaled counts curve overlain. Figure~\ref{fig:tiger_figure} (a) shows posterior probability of the number of changepoints over the 100 runs of the sampler. It can be seen that the sampler performs consistently, giving similar results over the 100 runs. Figure~\ref{fig:tiger_figure} (b) shows a histogram for the sampled values of $p$ from the last run. Posterior support for $p$ is highest over the range $[0,0.1]$.

%\noindent{\it Filtering recursions}

\begin{figure}
\begin{center}
\includegraphics[width=65mm,height=65mm]{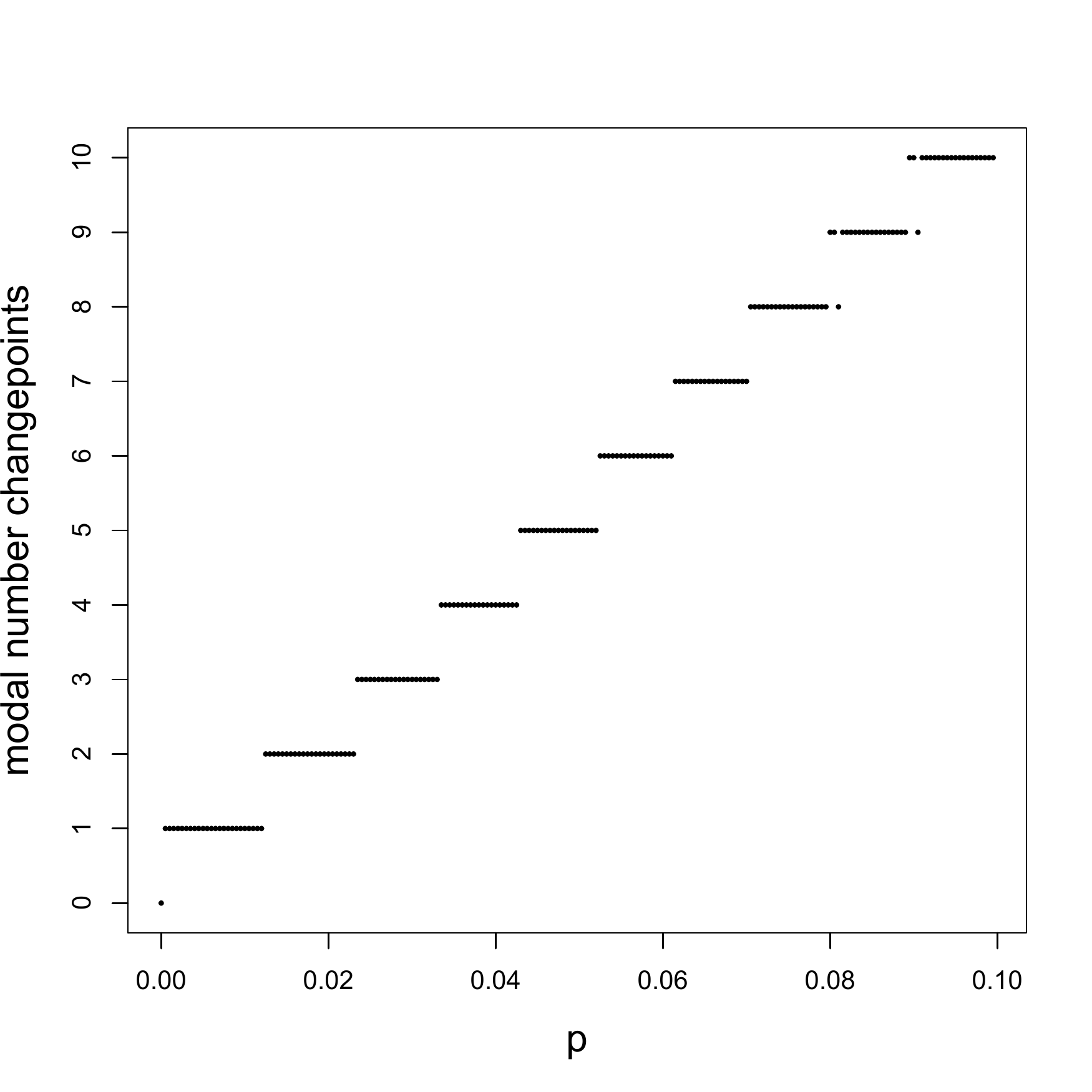}
\end{center}
\caption{Streakiness data: Modal number of changepoints from a filtering recursions analysis over a range of values of $p$.} \label{fig:sensitivity_tiger_woods}
\end{figure}

For the filtering recursions analysis~\cite{Fearnhead06}, the recursions of Section~\ref{sec:recursions} were computed for $p\in[0,0.1]$ following the analysis above. A sample of size 100,000 changepoints was generated and the posterior of the number of changepoints was computed for each value of $p$. The modal number of changepoints was recorded from this for each value of $p$ and is shown in Figure~\ref{fig:sensitivity_tiger_woods}. It is clear that the number of changepoints inferred in the filtering recursions analysis is very sensitive to the value of $p$ for this data. It is questionable whether such an analysis would be useful for a practitioner since it is unclear how one could objectively choose $p$ in this situation. Certainly an exploratory analysis would be necessary before choosing the value of $p$ to compute the filtering recursions. One suggestion is to use the sampler proposed here for an exploratory analysis of the posterior allowing for uncertainty in the specification of $p$. The MCMC sampler simulation study suggests that two changepoints is most likely although there is relatively strong support for up to five changepoints. In this case, specification of one value of $p$ to generate samples of changpoints will not fully explore uncertainty in the posterior. As before, the output of the MCMC sampler shown from Figure~\ref{fig:tiger_woods} (b) shows that one change is clearly identified, but that there is considerable uncertainty in the other positions, hence the support for up to five changepoints.

\section{Gaussian changepoint models} \label{sec:well_log_data}

Gaussian changepoint models are widely used and studied. Models can include those with changing mean and/or variance across segments. The model assumed for the purposes of the example here is piecewise constant, where data in any segment is Gaussian distributed. Segments share a common error variance. Data point $y_i$ in segment $j$ is assumed to arise independently from a
$
\mbox{N}(\mu_j,\sigma^2)
$
distribution. The segment means $\mu_j$ are assumed to arise from a Gaussian distribution with mean $\mu_0$ and variance $\nu^2 \sigma^2$, {\it a priori}. Denote $\gamma=(\sigma^2,\mu_0,\nu^2)$. Segment length is assumed to have a geometric distribution with parameter $p$. This gives the log posterior (up to a constant) as
\begin{eqnarray*}
\log \pi(k,z|y,p,\gamma) &=&  -(k+1) \log \nu - (n+k+1)\log \sigma +(n-k-1)\log(1-p) +k \log p  \\
& & -\frac{1}{2} \sum_{j=1}^{k+1} \left\{ \log\left(\tau_j - \tau_{j-1} + \frac{1}{\nu^2}\right)  - \frac{1}{\sigma^2}\left( ss_j + \frac{\mu_0^2}{\nu^2} - \frac{\left(s_j + \frac{\mu_0}{\nu^2} \right)^2}{\tau_j - \tau_{j-1} + \frac{1}{\nu^2}}\right) \right\},
\end{eqnarray*}
where $ss_j = \sum_{i=\tau_{j-1}+1}^{\tau_j}y_i^2$ and $s_j=\sum_{i=\tau_{j-1}+1}^{\tau_j}y_i$. Details of this calculation are given in the Appendix.
\newline

\subsection*{Application to Well-log data}
%\noindent{\it Application to Well-log data}

\begin{figure}
\begin{center}
$\begin{array}{c}
\includegraphics[width=16.5cm,height=6cm]{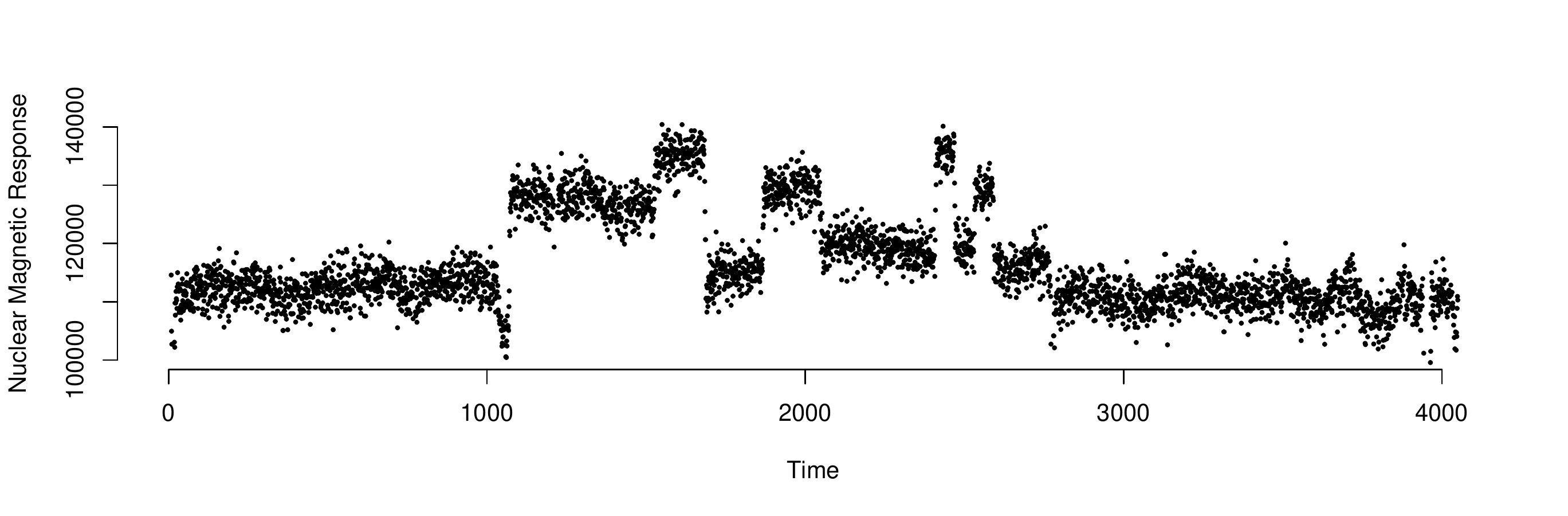}\\
\includegraphics[width=16.5cm,height=6cm]{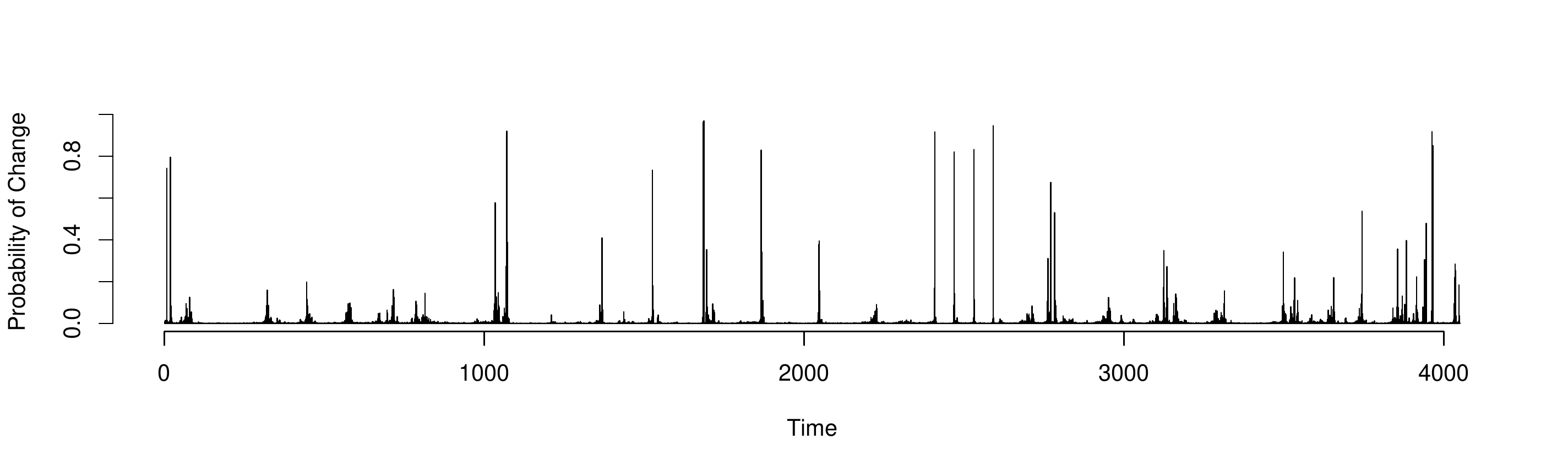}
\end{array}
$
\end{center}
\caption{Top: Well-log data. Bottom: Posterior probability of a changepoint in any position from 100,000 samples using the sampler with hyperpriors. } \label{fig:well_log_data}
\end{figure}

 The Well-log data (\citeasnoun{ORuanaidh96}) records measurements of nuclear-magnetic response of underground rocks obtained by lowering a probe into a bore-hole. The probe records the response at regular points in time. As well as~\citeasnoun{Fearnhead06} this data was also analyzed in~\citeasnoun{Fearnhead03}. The data consists of 4050 measurements, some of which are outliers and were removed before analysis. The data are shown in Figure~\ref{fig:well_log_data}.

The purpose of this example is to demonstrate how results from an analysis with filtering recursions may be sensitive to the choice of hyperparameters $\gamma$ and how a short run of the sampler could possibly provide good starting values. It is possible to fit a more elaborate state space model to the Well-log data, however, this is not considered here.

\begin{figure}
\begin{center}
$
\begin{array}{ccc}
\includegraphics[width=60mm]{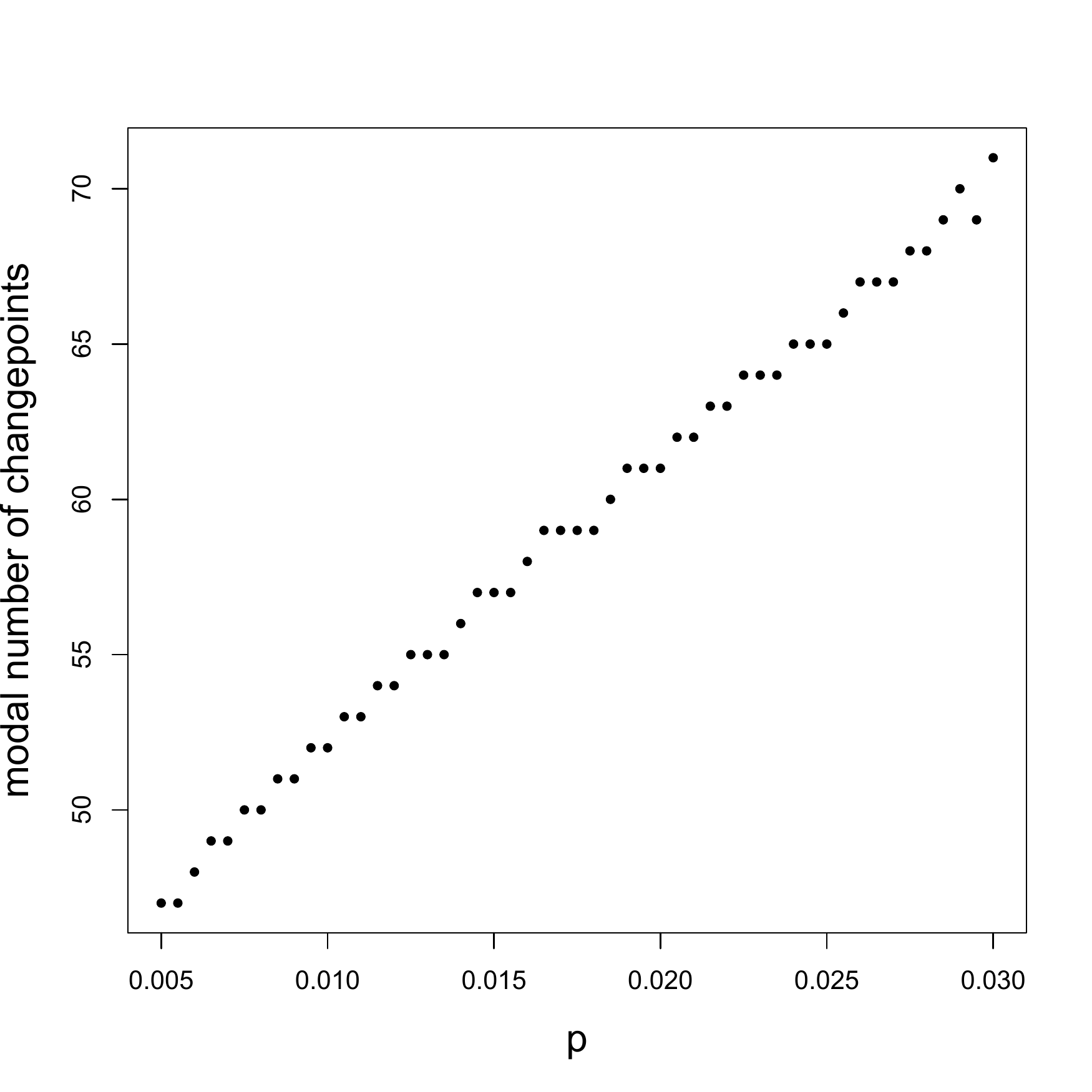} & 
\includegraphics[width=60mm]{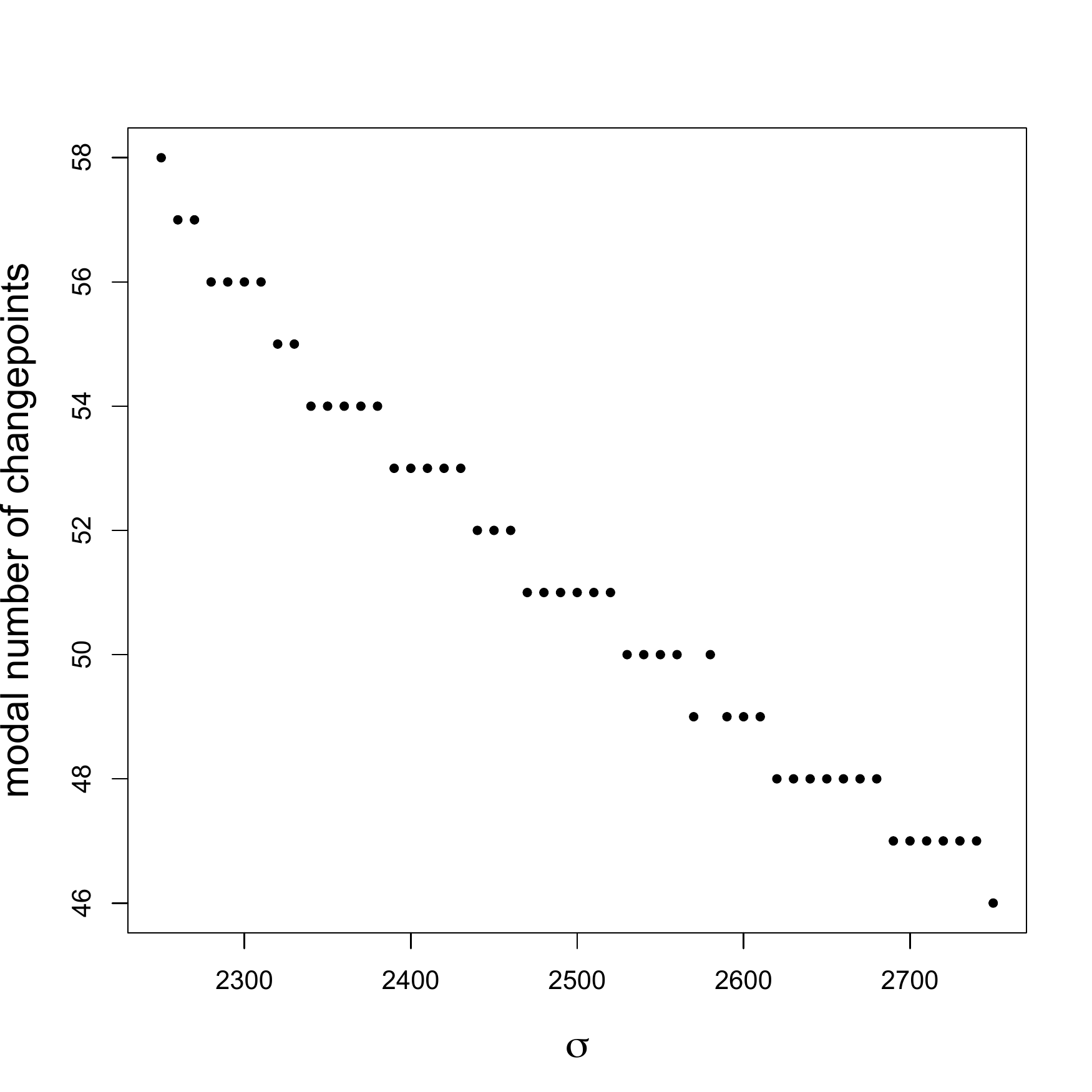} 
%&\includegraphics[width=60mm]{well_log_tau_sensitivity}
\end{array}
$
\end{center}
\caption{Well-log data: Modal number of changepoints for a filtering recursions analysis of the Well-log data for Experiment 1 and Experiment 2. Experiment 1 varies $p$ (left) and Experiment 2 varies $\sigma$ (right)} \label{fig:sensitivity_p_and_sigma}
\end{figure}

~\citeasnoun{Fearnhead06} chose the values $p = 0.013, \sigma = 2,330, \nu = 4.3, \mu_0 = 115,000$ when analyzing the Well-log data in the section on inclusion of hyperpriors. Two simple experiments were performed here to investigate sensitivity of the posterior distribution to prior specification. One of $p$ (Experiment 1) or $\sigma$ (Experiment 2) was varied over a grid on a small range keeping all other hyperparameter values fixed (details in Table~\ref{tab:sensitivity_exp}). The recursions of Section~\ref{sec:recursions} were computed for each value on the grid and a sample of size 100,000 was generated from the posterior of the changepoints. The empirical posterior distribution of the number of changepoints was computed for each of these samples and the modal number of changepoints recorded. The results are summarized in Figure~\ref{fig:sensitivity_p_and_sigma}. It can be seen that the modal value of the posterior number of changepoints is sensitive to the values of both $p$ and $\sigma$. Thus choosing these values, {\it a priori}, places the posterior mass $\pi(k,z|y,p,\gamma)$ in the area determined by $p$ and $\sigma$ and may not correctly represent the true posterior over all $p,\sigma$.

For the Well-log data it would seem most sensible to carry out an analysis with inclusion of hyperpriors on $p,\sigma$ and $\mu_0$ using the scheme outlined in Section~\ref{sec:priors}. The hyperpriors used are $\pi(p) \propto 1$, $\pi(\mu_0) \propto 1$, $\pi(\nu)\propto 1/\nu$, $\pi(\sigma)\propto 1/\sigma$. The bottom of Figure~\ref{fig:well_log_data} shows the posterior probability of a change output from an algorithm run for 10,000 burn-in and 100,000 subsquent iterations using a random walk update for changepoint positions. Ergodic mean estimators of the hyperparameters were $\hat{\sigma} = 2360,\hat{p}=0.014,\hat{\nu}=3.99,\hat{\mu}_0 = 113771.0$. This took about 10 sec on a 2.5GHz processor with very diffuse starting values. This Gaussian model infers many changepoints as it picks up small changes in the mean and thus performs well for this data.

\begin{figure}
\begin{center}
$
\begin{array}{cc}
\includegraphics[width=60mm]{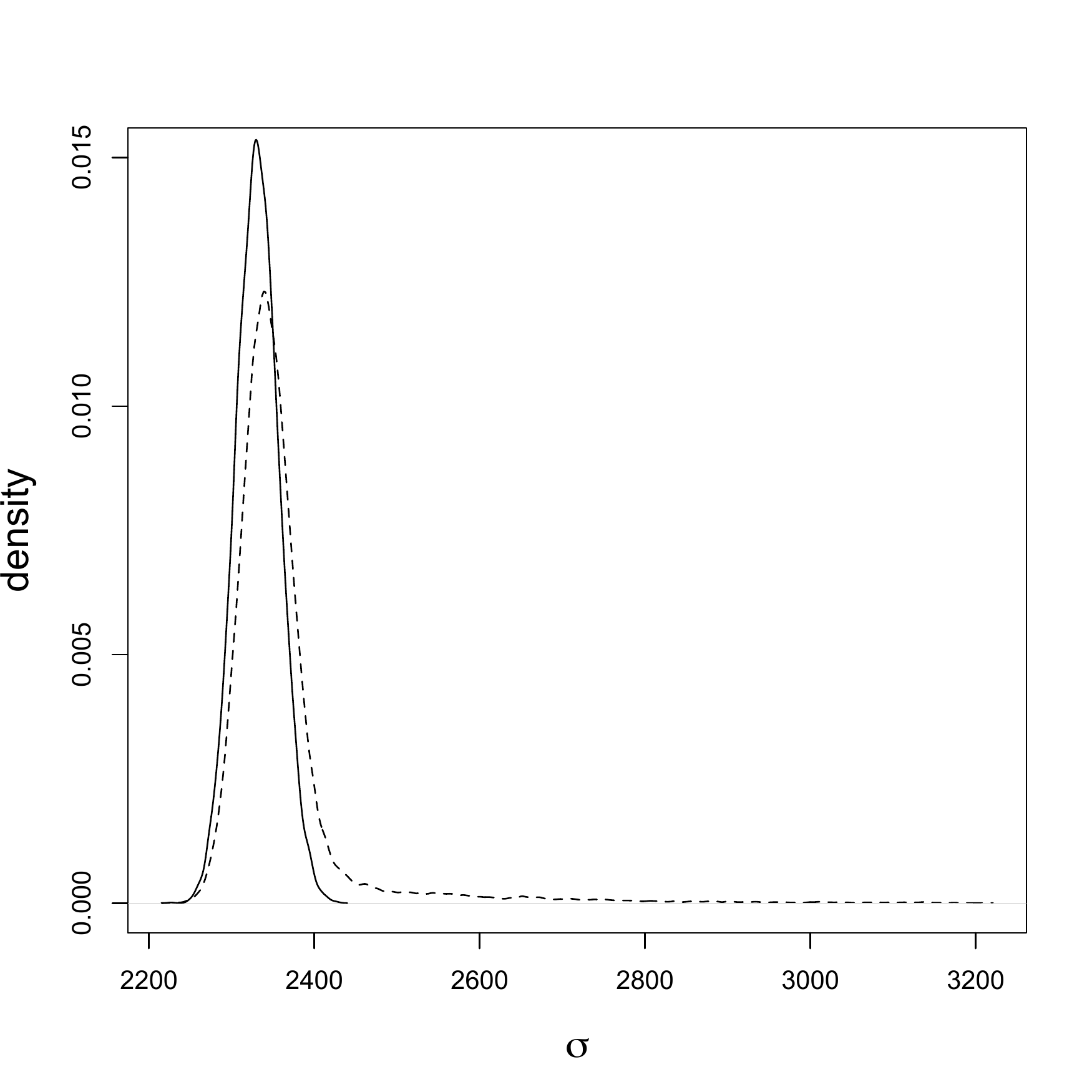} &
\includegraphics[width=60mm]{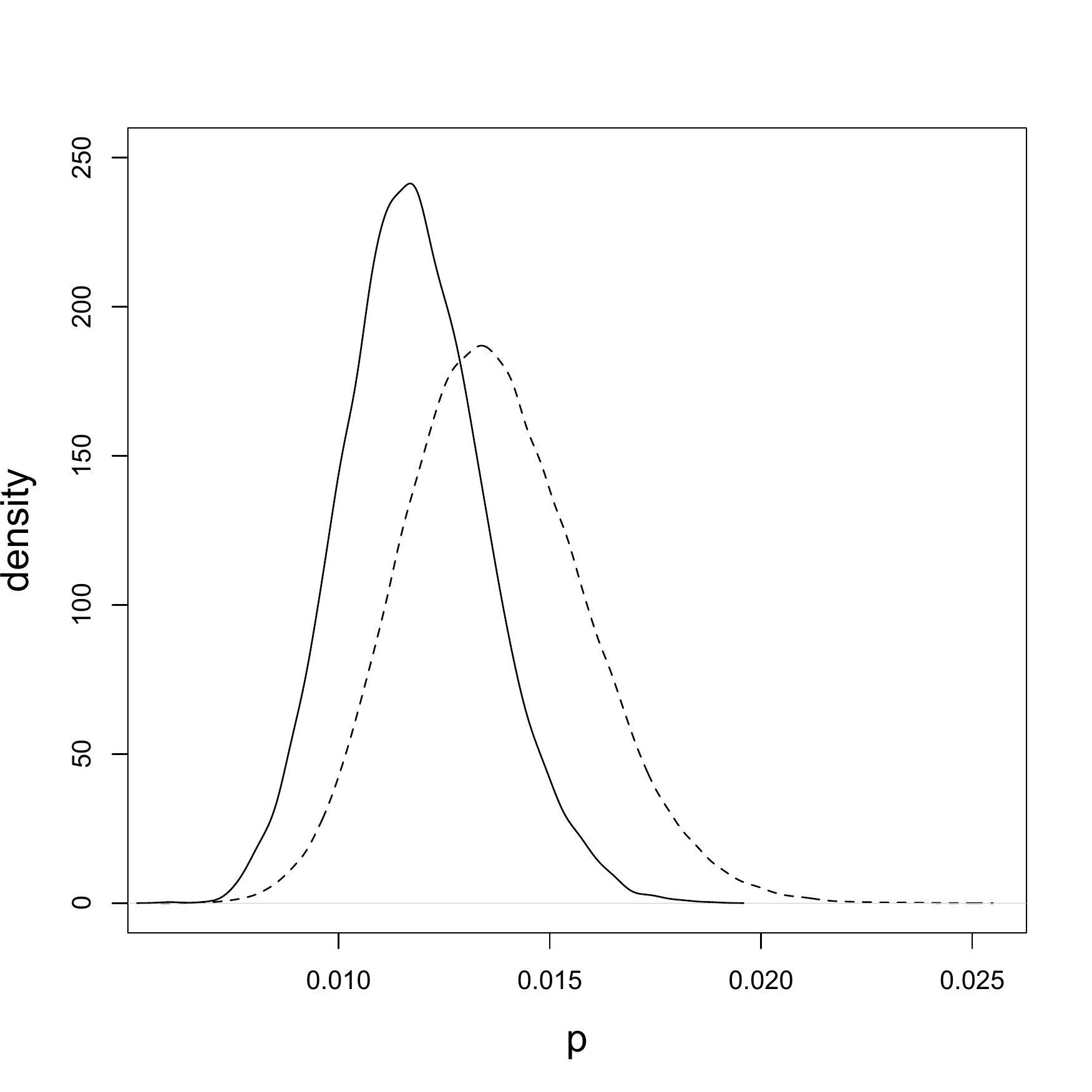} \\
\includegraphics[width=60mm]{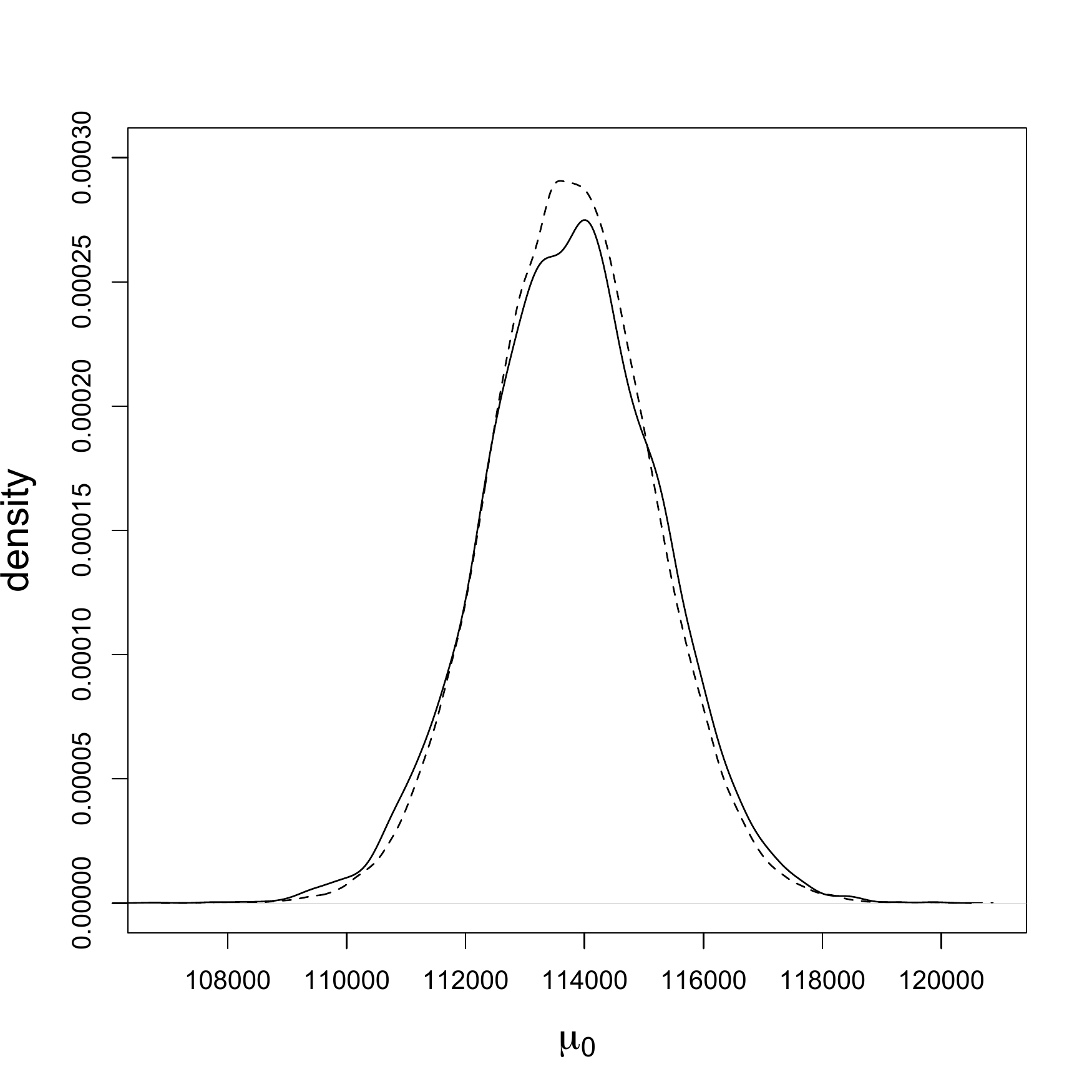} &
\includegraphics[width=60mm]{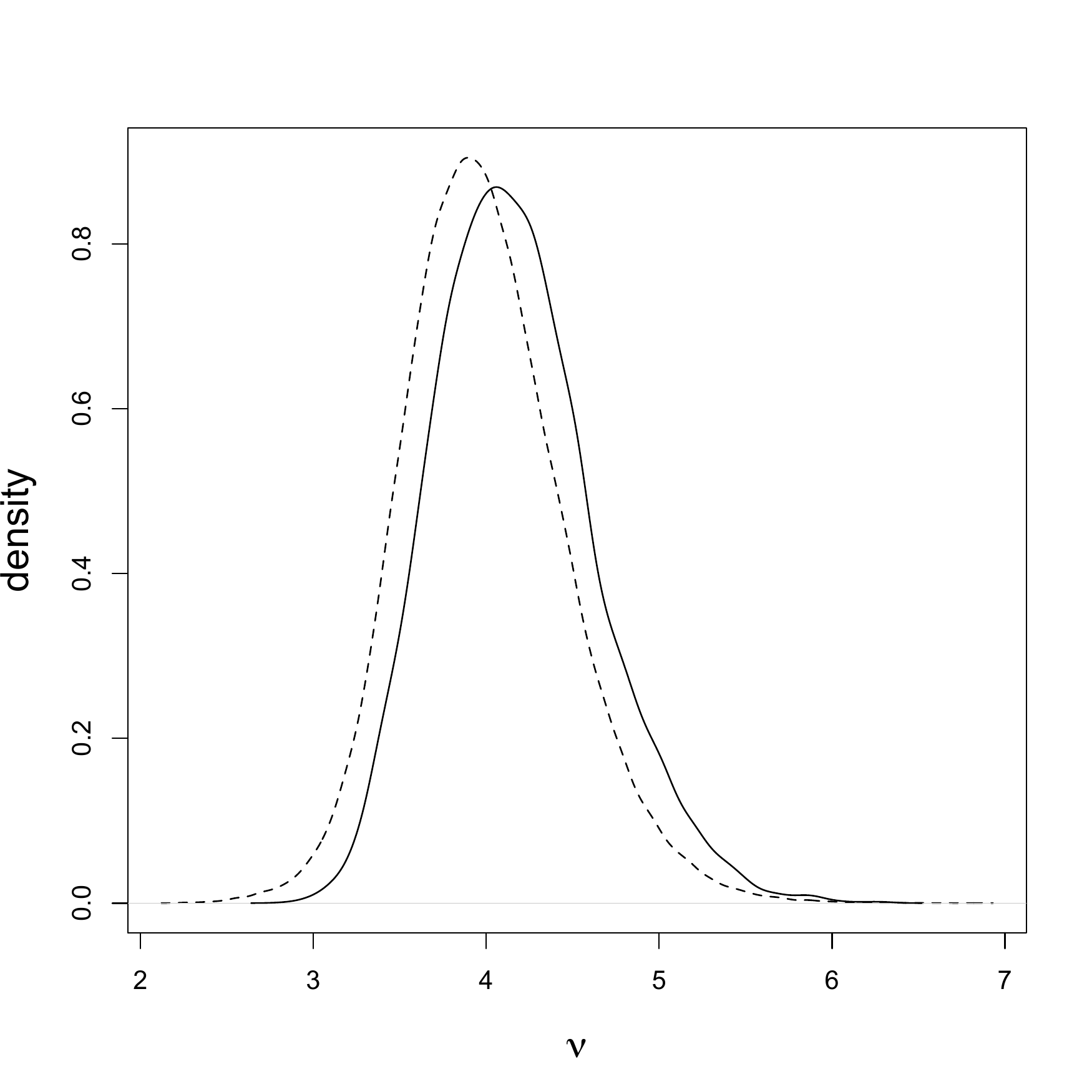}
\end{array}
$
\end{center}
\caption{Well-log data: Comparison of long run of sampler to MCMC scheme with independent proposals from filtering recursions. Dashed lines give the density from the MCMC sampler output and solid lines give the density output from analysis using the independent proposal scheme suggested in Fearnhead (2006). } \label{fig:comparison_independence_proposal}
\end{figure}

A long run of the sampler was implemented so as to obtain a near independent sample ($1.8\times10^7$ iterations taking every $1,800^{\mbox{th}}$ sample; estimated integrated autocorrelation time of the number of changepoints $\approx 1$) of size 10,000 from the posterior distribution of changepoints and hyperparameters. This was compared with results from the independence proposal suggested by~\citeasnoun{Fearnhead06}. In the independence proposal MCMC scheme suggested in~\citeasnoun{Fearnhead06}, a sample of changepoints is generated using filtering recursions conditional on $p = 0.013,\sigma = 2,330,\mu_0 = 115,000,\nu = 4.3$. This sample is then used for an independence proposal and hyperparmeters are updated in the same way as done here. Figure~\ref{fig:comparison_independence_proposal} shows kernel density estimates constructed from samples of the hyperparameters for the sampler (dashed line) and independence proposal (solid line). It can be seen that there is a slight discrepancy in that the independence proposal leads to more peaked densities. 

In our implementation an independence proposal based on a sample of size 10,000 was used. This updating scheme for hyperparameters and changepoints was then run for 50,000 iterations. Although the acceptance rate for moving between different changepoint configurations was high, the independence proposal distribution was highly degenerate. Only ten unique changepoint configurations were sampled in the 50,000 iterations of the MCMC scheme. For other datasets where less information is available to choose the hyperparameters to generate the independence proposal, it is possible that this could lead to highly biased sampling from the hyperpriors.

In the sense of hyperprior incorporation and full exploration of the posterior distribution the MCMC sampler proposed performs better than the independence proposal. However, generating independent samples may be more costly in large datasets with many changepoints. Nonetheless, it is clear that the inclusion of hyperpriors circumvents the sensitivity of posterior distribution of the changepoints to specification of the hyperparameters. This is a main advantage of the approach proposed here and makes the detection of changepoints more automatic.

\begin{table}
\begin{center}
\begin{tabular}{|l|c|c|}
\hline
Recursion Sensitivity &Fixed & Varied \\
\hline
Experiment 1 &$\sigma = 2,330, \nu = 4.3, \mu_0 = 115,000$ & $p \in [0.005,0.03]$\\
Experiment 2&  $p = 0.013, \nu = 4.3, \mu_0 = 115,000$ & $\sigma \in [2250,2750]$\\
%Experiment 3    &$p = 0.013,\sigma = 2,330,\mu_0 = 115,000$ & $\nu \in [3.5,5.0]$\\
 \hline
 \end{tabular}
\end{center}
\caption{Well-log data: Experiments to investigate sensitivity of results of filtering recursions to prior specification} \label{tab:sensitivity_exp}
\end{table}

\section{Discussion}

This paper has presented an MCMC method to perform retrospective inference for changepoint model which are collapsable. The multiple changepoint problem is rephrased as a stochastic model search over a large models space, with the Bayes factors for competing models appearing in the acceptance probabilities for the MCMC sampling scheme.

The performance of the sampler was verified for the benchmark coal mining disasters data. Application of the sampler to a streakiness dataset from sports revealed that posteriors for the number of changepoints can be diffuse. It was demonstrated that prior specification on the duration of segments plays a crucial role in the analysis of the models considered. Incorporation of hyperpriors to account for this revealed features of the posterior that would be missed by a popular filtering recursions analysis for changepoints. Application to the Well-log data further highlighted sensitivity of analysis by filtering recursions to prior specification. It was shown that output from a short run of our sampler can be used to give good values of the hyperparameters for this prior specification.

In conclusion, the sampling scheme presented is shown to work well and can provide further insight and account for prior uncertainty in some difficult situations. It can be used as a useful exploratory tool or for a full analysis. Computer code implementing the sampler written in C may be downloaded from \texttt{www.ucd.ie/statdept/jwyse}.

\section*{Appendix}

\subsection*{Calculations for the coal-mining example}

Given a segment $y_{s:t}$, each $y_i\sim_{\mbox{\tiny iid}} \mbox{Poisson}(\mu)$. Here $\mu$ is the height of the step function that gives the intensity of the process between times $s$ and $t$. Assume the prior for $\mu$ is $\mbox{Gamma}(\rho,\lambda)$ where $\gamma = (\rho,\lambda)$. The marginal likelihood for the segment is then
\begin{eqnarray*}
\pi(y_{s:t}|\gamma) &=& \int_{0}^{\infty} \frac{\lambda^{\rho}}{\Gamma\{\rho\}} \mu^{\rho-1}\exp\{-\lambda \mu\} \prod_{i=s}^t \frac{\mu^{y_i}}{y_i!} \exp\{-\mu\} \, \mbox{d}\mu \\
 & = & \frac{\lambda^{\rho}}{\Gamma\{\rho\}} \int_{0}^{\infty} \frac{1}{F_{s:t}}\mu^{S_{s:t}+\rho-1} \exp\{-(t-s+\lambda+1)\mu\} \, \mbox{d}\mu
\end{eqnarray*}
where $F_{s:t} = \prod_{i=s}^t y_i!$ and $S_{s:t} = \sum_{i=s}^t y_i$. Completing the integral of the Gamma density gives
\[
\pi(y_{s:t}|\gamma)  =  \frac{\lambda^{\rho}}{\Gamma\{\rho\}} \frac{1}{F_{s:t}} \frac{\Gamma\{S_{s:t}+\rho\}}{(t-s+\lambda+1)^{S_{s:t}+\rho}}
\]

\subsection*{Calculations for the streakiness example}
Within a segment $y_{s:t}$, $y_i \sim_{\mbox{\tiny iid}} \mbox{Bernoulli}(\phi)$. Taking a $\mbox{Beta}(\alpha,\beta)$ prior on $\phi$, the marginal likelihood is obtained from
\[
\pi(y_{s:t}|\gamma) = \int_{0}^1 \frac{\Gamma\{\alpha+\beta\}}{\Gamma\{\alpha\}\Gamma\{\beta\}}\phi^{\alpha-1} (1-\phi)^{\beta-1} \prod_{i=s}^t \phi^{y_i}(1-\phi)^{1-y_i} \, \mbox{d} \phi
\]
where $\gamma=(\alpha,\beta)$. This reduces to
\[
\pi(y_{s:t}|\gamma) = \frac{\Gamma\{\alpha+\beta\}}{\Gamma\{\alpha\}\Gamma\{\beta\}} \int_{0}^1\phi^{S_{s:t} + \alpha-1} (1-\phi)^{t-s-S_{s:t}+\beta} \, \mbox{d} \phi.
\]
where $S_{s:t} = \sum_{i=s}^t y_i$. Completing the Beta integral gives
\[
\pi(y_{s:t}|\gamma) = \frac{\Gamma\{\alpha+\beta\}}{\Gamma\{\alpha\}\Gamma\{\beta\}} \frac{\Gamma\{S_{s:t}+\alpha\}\Gamma\{t-s+1-S_{s:t}+\beta\}}{\Gamma\{t-s+1+\alpha+\beta\}}.
\]

\subsection*{Calculations for Gaussian changepoint model}

The model for all the data may be written hierarchically as
\begin{eqnarray*}
\pi(k,z,\theta|y,p,\gamma) & \propto &  \pi(z|k,p) \pi(\theta|k,z,\sigma,\mu_0) \pi(y|k,z,\theta) \\
&\propto & p^k (1-p)^{n-k-1} \prod_{j=1}^{k+1} \frac{1}{\nu\sigma\sqrt{2\pi}} \exp\left\{-\frac{1}{2\nu^2\sigma^2}(\mu_j-\mu_0)^2\right\} \\
&& \qquad \qquad \times \prod_{i=\tau_{j-1}+1}^{\tau_j} \frac{1}{\sigma \sqrt{2 \pi}} \exp\left\{-\frac{1}{2\sigma^2} (y_i - \mu_j)^2 \right\} \\
& =  &  \frac{(2 \pi)^{-(n+k+1)/2}}{\nu^{k+1} \sigma^{n+k+1}} p^k (1-p)^{n-k-1} \\ && 
\prod_{j=1}^{k+1} \exp\left\{-\frac{1}{2\sigma^2} \left[\left(\tau_j - \tau_{j-1} + \frac{1}{\nu^2} \right)\mu_j^2-2\left(s_j+\frac{\mu_0}{\nu^2}\right) \mu_j + ss_j + \frac{\mu_0^2}{\nu^2}\right]\right\}.
\end{eqnarray*}
Completing the square on $\mu_j$ and then performing integration of $\mu_j$ over $(-\infty,\infty)$ gives the required posterior.
\begin{eqnarray*}
\pi(k,z,\theta|y,p,\gamma) & \propto & \frac{(2 \pi)^{-n/2}}{\nu^{k+1} \sigma^{n}} p^k (1-p)^{n-k-1}\\ & &\prod_{j=1}^{k+1}  \left(\tau_j - \tau_{j-1} + \frac{1}{\nu^2} \right)^{1/2} \exp\left\{-\frac{1}{2\sigma^2} \left( ss_j + \frac{\mu_0^2}{\nu^2} - \frac{\left(s_j + \frac{\mu_0}{\nu^2} \right)^2}{\tau_j - \tau_{j-1} + \frac{1}{\nu^2}}\right)\right\}
\end{eqnarray*}

\bibliography{paper_bibliography}
\end{document}